\documentclass[twocolumn,showpacs,preprintnumbers,amsmath,amssymb,showkeys]{revtex4}

\usepackage{graphicx}% Include figure files
\usepackage{dcolumn}% Align table columns on decimal point
\usepackage{bm}% bold math
\usepackage{color}

\begin{document}

\preprint{APS/123-QED}

\title{Anomalous pH-gradient in Ampholyte Solution}% Force line breaks with \\

\author{L.\;V.~Sakharova}
\email{L_Sakharova@mail.ru}
\affiliation{%
Naval State Academy, 344006, Rostov-on-Don, Russia
}%
\author{V.\;A.~Vladimirov}
 \email{vv500@york.ac.uk}
\affiliation{%
Department of Mathematics, York University, York, YO10 5DD, UK
}%
\author{M.\;Yu.~Zhukov}
\email{zhuk@ns.math.rsu.ru}
\affiliation{%
Department of Mathematics, Mechanics and Computer Science, Southern
Federal University, 344090, Rostov-on-Don, Russia
}%

\date{\today}% It is always \today, today,

\begin{abstract}
A mathematical model describing a steady pH-gradient in the solution of ampholytes in water has been studied with the
use of analytical, asymptotic, and numerical methods. We show that at the large values of an electric current a
concentration distribution takes the form of a piecewise constant function that is drastically different from a
classical Gaussian form. The correspondent pH-gradient takes a stepwise form, instead of being a linear function. A
discovered anomalous pH-gradient can crucially affect the understanding of an isoelectric focusing process.
\end{abstract}

%Use showkeys class option if keyword
\maketitle

\section{Introduction}

The isoelectric focusing (IEF) is a classical method of a mixture separation that is widely used in  medicine and
biology. There is an enormous number of publications devoted to its theory and applications. We mention here only
fundamental papers
\cite{Rilbe76,Rilbe76b,Vesterberg,Vesterberg1,Haglund71}, the monographs considering various theoretical and practical
aspects of the IEF \cite{Righetti83,Righetti90,ZhRStoy2001}, and papers
\cite{MosherBierRighetti,SavillePalusinski1,Cann,TsaiMosherBier, BusasHjelmelandChrambach,MosherSavilleThorman,
MosherThormanGrahamBier,NguyenChrambach,MosherThorman, ThormannPawliszyn,BabZhukYudE,Zhukov2005}
that contain mathematical models and computer simulations that are close to the topic of our paper. There are also a
number of papers that study the related problems in space biotechnology
\cite{ZhZiva94,ZhZiva95,ZhPetr97MZhG,ZhSazonovDifUr97, BZhMyshkis,Bello,Polezaev-3}.

An important part of the IEF is the creation of stable pH-gradients that are used for the setting up of a focusing
process. Carrier ampholytes have been in a wide practical use for the creation of pH-gradients starting from pioneering
papers \cite{Rilbe76,Rilbe76b,Vesterberg,Vesterberg1,Haglund71} (see also \cite{Righetti83,MosherThorman}). The use of
immobilized pH-gradients for the same purpose has been proposed and introduced by
\cite{Righetti90,MosherBierRighetti}, while the use of borate-polyol systems for the creation of pH-gradients is
considered in \cite{Troickii}. The mathematical models describing the formation of stable pH-gradients with
the use of carrier ampholytes are given in
\cite{BabZhukYudE,MosherSavilleThorman,MosherThorman,TsaiMosherBier,Zhukov2005}. In particular,
\cite{BabZhukYudE,Zhukov2005} introduce the notion of mixtures with the infinite number of
components (the infinite-component mixtures) where the authors formally replace the discrete number $k$ in the
concentration $c_k$ onto the continuous parameter $s$; in other words the discrete set $c_k$, $k=1,\dots,n$ is replaced
by the distribution function $c(s)$. The first and simplest mathematical models describing the creation of linear (or
almost linear) pH-gradients with the use of carrier ampholytes were given in \cite{Haglund71,Rilbe76b}. More advanced
approaches that employ the governing differential equations of continuous media, mathematical modeling, and
computations are given in \cite{MosherSavilleThorman,MosherThorman}. These papers show that an ampholyte concentration
distribution should be close to Gaussian's one and a pH-gradient should be linear. Similar results with the use of
models for infinite-component mixtures were obtained in \cite{BabZhukYudE}.

The statement that a concentration distribution possesses a Gaussian shape is currently accepted as a key classical
result. Moreover, it is widely assumed that the increase of a potential difference (or an electric current) stabilizes
a linear pH-gradient and intensifies a focusing process. On the contrary, the main results of our paper show that the
increase of an electric current (starting from some threshold value) causes an anomalous steady stepwise pH-gradient
and a piecewise-constant concentration distribution. In addition we present explicit analytical expressions for a pH
distribution and for an electric conductivity as the functions of ampholyte concentrations. The obtained asymptotic
expressions for the concentrations of ampholytes allow us to carry out the calculations of stepwise pH-gradients.

\section{Mathematical Model}\label{i:1.0}

Well-known governing equations describing IEF in the aqueous  solution of ampholyts in dimensionless variables are
\cite{BabZhukYudE,Zhukov2005,MosherSavilleThorman}:
\begin{eqnarray}
&& \frac{\partial c_k}{\partial t}+\frac{\partial\, i_k}{\partial x}=0, \quad k=1,\dots,n,
    \label{vvz:1}
\end{eqnarray}
\begin{eqnarray}
&& i_k =
   -\varepsilon\gamma_{k}\frac{\partial c_k}{\partial x}+\gamma_{k} e_{k}(\psi)c_k E,
    \label{vvz:2}
\end{eqnarray}
\begin{eqnarray}
 j&=&\sum_{k=1}^{n}\gamma_{k}\left(-\varepsilon
\frac{\partial \left(e_{k}(\psi)c_{k}\right)}{\partial x}+
  \sigma_{k}(\psi)c_{k}E\right)\nonumber\\
  &&{}+2K_w \gamma_{0} \left(E-\varepsilon
  \frac{\partial \psi}{\partial x}\right)\cosh(\psi-\psi_0),
    \label{vvz:3}
\end{eqnarray}
\begin{eqnarray}
&& \sum_{k=1}^{n}e_{k}(\psi)c_{k}+2K_w \sinh\psi=0,
    \label{vvz:4}
\end{eqnarray}
where
\begin{eqnarray}
&& \gamma_{0}=\sqrt{\gamma_{_H} \gamma_{_{OH}}}, \quad
 \psi_0=\frac12\ln \frac{\gamma_{_{OH}}}{\gamma_{_H}},
     \label{vvz:5}
\end{eqnarray}
\begin{eqnarray}
&&   h=K_we^{\psi}, \quad {\textrm pH}=-\lg h=-\lg K_w-\psi\lg e.
     \label{vvz:6}
\end{eqnarray}
Here $c_k(x,t)$ are molar ampholyte concentrations,
      $\psi(x,t)$ is a function, linearly connected to pH of a solution,
      $E(x,t)$ is the strength of an electric field,
      $i_k$ are the densities of molar ampholyte fluxes,
      $j=\text{\rm const}$ is the density of an electric current (it can also depend on time  $j=j(t)$),
      $h$ is a hydrogen ion concentration,
      $\varepsilon$ is a diffusion parameter,
      $\gamma_{k}>0$ are characteristic mobilities of components in an electric field,
      $\varepsilon \gamma_k>0$ are diffusion coefficients,
      $e_{k}(\psi)$, $\sigma_{k}(\psi)$ are the molar charges and the molar conductivities of ampholytes,
      $\gamma_k e_k(\psi)$ are the electrophoretic  mobilities of ampholytes,
      $\gamma_H$, $\gamma_{OH}$ are the mobilities of hydrogen  ${\textrm H}^{+}$ and hydroxyl $\textrm{OH}^{-}$ ions,
      $\gamma_0$ is the effective mobility of water ions,
      $K_w$ is the autodissociation constant for water,
      $\psi_0$ is the value of $\psi$ when water conductivity is minimal.

The system of equations (\ref{vvz:1})--(\ref{vvz:6}) describes the mass transport under the action of an electric
field  in a chemically active medium. Chemical reactions in the medium are assumed to be very fast (being completed
almost instantly). The functions $e_{k}(\psi)$, $\sigma_{k}(\psi)$ are given; in each particular case they can be
defined from the chemical kinetic equations (see below). The \emph{algebraic} equation (\ref{vvz:4}) represents the
condition of mixture electroneutrality; it allows us to find $\psi$. In fact, this equation describes the instant
control of medium properties (electrophoretic mobilities and molar conductivities) by the function $\psi$ (that is
linked to the concentration of hydrogen ions). Such situation is typical only for a weak aqueous  solution;  in more
general cases the values $e_{k}$, $\sigma_{k}$ can depend on $c_k$. The term $2K_w \gamma_{0} \left(E-\varepsilon
\partial\psi/\partial x\right)\cosh(\psi-\psi_0)$ in (\ref{vvz:3}) describes the contribution of water ions into
mixture conductivity, while the term $2K_w \sinh\psi$ in  (\ref{vvz:4}) corresponds to the contribution of these ions
into a mixture molar charge. These terms should be included into the governing equations if the autodissociation reaction
is taken into account:
\begin{eqnarray*}
  && \textrm{H}_2\textrm{O} {\,\overset{k^+}{{\underset{k^-}{\rightleftharpoons}}}\,}
     \textrm{H}^{+} + \textrm{OH}^{-}.
\end{eqnarray*}
In this case the  concentrations of ions $H^+$ and $OH^-$ are reflated as
\begin{eqnarray*}
  && h=[\textrm{H}^+],  \quad  [\textrm{OH}^-]=\frac{K^2_w}{[\textrm{H}^+]}=\frac{K^2_w}{h},
\end{eqnarray*}
where $K^2_w$ is an autodissociation constant for water, square brackets [~$\cdot$~] denote the values of concentrations,
as it is accepted in chemistry.

Let us note that the maximal values of concentrations $c_k$ and the values $\gamma_0$, $\psi_0$, $\gamma_k$, $\psi_k$,
$k=1,\dots,n$ are all of the order $O(1)$, while the parameters $\varepsilon$ and $K_w$ are small. Notice,
that we denote an autodissociation constant for water as $K_w^2$, not $K_w$; hence $K_w$ represents a dissociation
constant for water, its value in dimensional variables is $K_w=10^{-7}\,\textrm{mol/l}$. In addition, instead of a
conventionally used function pH (measure of the acidity or alkalinity of a solution) we use function $\psi$ (linearly connected with pH) which is better adapted to our
mathematical model (see (\ref{vvz:6})) \footnote{One can consider this step as being too radical. However the
introduction of new notations aimed to simplify mathematical expressions belong to the highest achievements in
sciences. For instance, the timely replacement of Latin terms for `plus', `minus', and `equals' by the symbols $+$,
$-$, and $=$ as well as the replacement of Roman figures for numbers by Arab digits initiated an extraordinary progress
in sciences.}.

In order to make the system (\ref{vvz:1})--(\ref{vvz:6}) complete, we have to prescribe the functions
$e_{k}(\psi)$, $\sigma_{k}(\psi)$ for ampholytes. Let us consider the solution of $n$ amphoteric substances
(ampholytes) that exhibit both acid and based properties (a typical representative is an amino acid). The dissociation
reactions are ($i=1,\dots,n$) \cite{Edsall,BabZhukYudE,Zhukov2005}:
\begin{eqnarray}
  && \textrm{H}^+ \textrm{R}
  \overset{B_i}{\rightleftharpoons}
  \textrm{R}_i^0 + \textrm{H}^+,
  \quad
  \textrm{R}^{0}_i
  \overset{A_i}{\rightleftharpoons}
     \textrm{R}^{-}_i + \textrm{H}^+.
  \label{vvz:7}
\end{eqnarray}
Here $\textrm{R}^{0}_i$ is a zwitterion (for instance an amino acid residue), $A_i$ and $B_i$ are dissociation constants for
acid ($\textrm{R}_i^-$) and based ($\textrm{H}^+\textrm{R}_i$) groups correspondingly. Notice, that even for polyampholytes  with supplementary
side groups a reaction scheme can be described by (\ref{vvz:7}), assuming the pH of a solution being close to the
isoelectric point of a substance \cite{ZhRStoy2001}. The analytical concentrations $c_i$, molar charges $e_i(\psi)$,
and molar conductivities $\sigma_i(\psi)$ for (\ref{vvz:7}) are:
\begin{eqnarray*}
  && c_i=[\textrm{H}^+\textrm{R}_i] + [\textrm{R}^0_i] +[\textrm{R}^{-}_i],
\end{eqnarray*}
\begin{eqnarray*}
  && e_{i}(\psi)=\frac{1}{\varphi_{i}(\psi)}\frac{d\varphi_{i}(\psi)}{d\psi},
 \quad
 \sigma_{i}(\psi)=\frac{1}{\varphi_{i}(\psi)}\frac{d^{\,2}\varphi_{i}(\psi)}{d\psi^2},
\end{eqnarray*}
\begin{eqnarray}
 && \varphi_{i}(\psi)=\delta_{i}+\cosh(\psi-\psi_{i}),
 \label{vvz:8}
\end{eqnarray}
\begin{eqnarray*}
  &&\psi_i=\frac12 \ln\frac{A_i B_i}{K_w^2}, \quad \delta_i=\frac12
 \left(\frac{B_i}{A_i}\right).
\end{eqnarray*}
Here $\psi_k$ is an isoelectric point (the value of $\psi$ when the molar charge $e_k$ is zero, \emph{i.e.}
$e_k(\psi_k)=0$), $\varphi_i$ and $\delta_i$ are auxiliary notations. Notice, that it is convenient \emph{not to write}
explicit expressions for  $e_k(\psi)$ and $\sigma_k(\psi)$ but to present them as the derivatives of functions
$\varphi_{i}(\psi)$.

\section{Formulation of Steady pH-gradient Problem}\label{i:2.0}

Let us take $\partial c_k/\partial t=0$ and introduce non-leak boundary conditions for the ampholytes at the walls of
an electrophoretic chamber ($0\leqslant x\leqslant 1$):
\begin{eqnarray}
&& i_k(0)=0, \quad i_k(1)=0, \quad k=1,\dots,n.
    \label{vvz:9}
\end{eqnarray}
The total molar concentration  of each ampholyte is:
\begin{eqnarray}
&& \int\limits_{0}^{1}c_{k}(x)dx=M_{k}, \quad k=1,\dots,n.
    \label{vvz:10}
\end{eqnarray}
The values $\gamma_0$, $\psi_0$, $\gamma_k$, $\psi_k$, $M_k$, $k=1,\dots,n$, $K_w$,  $\varepsilon$ are given. For the
sake of simplicity we consider only the IEF case with a given electric current density $j=\text{\rm const}$. The problem with a
given constant voltage can be solved similarly, however it is more tedious. The integration of equations
(\ref{vvz:1}) with boundary conditions (\ref{vvz:9}) yields $i_k=0$ that can be written as:
\begin{eqnarray}
&&  -\varepsilon \frac{dc_{k}}{dx}+ e_{k}(\psi)c_k E=0, \quad k=1,\dots,n,
    \label{vvz:11}
\end{eqnarray}
The remaining equations keep their form:
\begin{eqnarray}
 j&=&\sum_{k=1}^{n}\gamma_{k}\left(-\varepsilon
\frac{d \left(e_{k}(\psi)c_{k}\right)}{dx}+
  \sigma_{k}(\psi)c_{k}E\right)\nonumber\\
  &&{}+2K_w \gamma_{0} \left(E-\varepsilon
  \frac{d\psi}{dx}\right)\cosh(\psi-\psi_0),
    \label{vvz:12}
\end{eqnarray}
\begin{eqnarray}
&& \sum_{k=1}^{n}e_{k}(\psi)c_{k}+2K_w \sinh\psi=0.
    \label{vvz:13}
\end{eqnarray}
The system of equations (\ref{vvz:11})--(\ref{vvz:13}) with additional conditions (\ref{vvz:10}) allow to find the
concentrations $c_k(x)$, $\quad k=1,\dots,n$, the strengths of the electric field $E(x)$, and the function $\psi(x)$
(\emph{i.e.} $\textrm{pH}(x)$). Let us emphasize here that despite of the presence of derivative $d\psi/dx$  we do not need to
introduce any supplementary condition for $\psi$, since the function $\psi$ has been determined by the algebraic
equation (\ref{vvz:13}).

\section{The Transformation of the Problem}\label{i:3.0}

For the problem (\ref{vvz:10})--(\ref{vvz:13}) we have found the change of dependent variables that allows us to
obtain explicit analytical expressions for functions $\psi(x)$ and $E(x)$. We introduce a new `concentration' $a_k(x)$
and the `effective' strength of the electric field $E_0(x)$ as:
\begin{eqnarray*}
&& c_{k}(x)=a_{k}(x)\varphi_{k}(\psi(x)),
\end{eqnarray*}
\begin{eqnarray}
&& E_{0}(x)=E(x)-\varepsilon\frac{d\psi(x)}{dx}.
  \label{vvz:14}
\end{eqnarray}
In this case one can easily show that $\psi(x)$ and $E(x)$ are defined by the explicit expressions, provided that
functions $a_k(x)$ are known:
\begin{eqnarray*}
 &&\psi(x)=\frac{1}{2}\ln\frac{2K_w +
\sum\limits_{k=1}^{n}a_{k}(x)e^{\psi_{k}}}
  {2K_w+\sum\limits_{k=1}^{n}a_{k}(x)e^{ -\psi_{k}}},
\end{eqnarray*}
\begin{eqnarray}
 &&E(x)=E_0(x)+\varepsilon\frac{d\psi(x)}{dx}, \quad   E_{0}(x)=\frac{j}{\sigma(x)},
    \label{vvz:15}
\end{eqnarray}
where
\begin{eqnarray}
 \sigma(x)&=&\sum\limits_{k=1}^{n}\gamma_{k} a_{k}(x)
 \left(\varphi_{k}(\psi)\frac{d^2}{d\psi^2}\ln{\varphi_{k}(\psi)}
 \right)_{\psi=\psi(x)}
 \nonumber\\
  &&{}+2K_w\gamma_{0}\cosh(\psi(x)-\psi_{0}).
    \label{vvz:16}
\end{eqnarray}
In order to find $a_k(x)$ (taking into account the conditions (\ref{vvz:10}) and the changes of variables
(\ref{vvz:14})) we have the following system of $2n$ differential equations with $\psi$ and $\sigma$ defined by
(\ref{vvz:15}) and (\ref{vvz:16})
\begin{eqnarray}
&& \varepsilon_0 \frac{d}{dx}\ln a_{k}(x)=
   \frac{1}{\sigma(x)}
   \left.\frac{d}{d\psi}\ln\varphi_{k}(\psi)\right|_{\psi=\psi(x)},
  \label{vvz:17}
\end{eqnarray}
\begin{eqnarray}
&& \frac{d}{dx}m_k(x)=a_{k}\varphi_{k}(\psi(x)),\quad
k=1,\dots,n.
    \label{vvz:18}
\end{eqnarray}
\begin{eqnarray*}
&& \varepsilon_0=\frac{\varepsilon}{j}, \quad
\left(m_k(x)=\int\limits_{0}^{x}c_k(s)\,ds\right),
\end{eqnarray*}
with the boundary conditions:
\begin{eqnarray}
 && m_{k}(0)=0, \quad m_{k}(1)=M_{k}, \quad   k=1,\ldots ,n.
     \label{vvz:19}
\end{eqnarray}
The additional variables $m_k$ have the meaning of the total number of moles of an ampholyte in the interval $[0,x]$.

\section{The Main Asymptotic Term at $K_w\to 0$}\label{i:4.0}

The nonlinear boundary problem (\ref{vvz:17})--(\ref{vvz:19}) can be solved numerically with the use of a shooting
method, by the reduction of (\ref{vvz:17})--(\ref{vvz:19}) to Cauchy's problem with unknown initial data $a_k(0)$. To
find $a_k(0)$ one should solve $n$ nonlinear transcendent equations. It can be performed for example by the
Newton-Raphson method that requires good initial approximations. An initial approximation for small $\varepsilon_0$ and
$K_w$ can be obtained with the use of asymptotic formulas. The analysis of equations (\ref{vvz:17}) show that their
right side parts at $\varepsilon_0 \to 0$ are proportional to $d\varphi_{k}(\psi)/d\psi\sim\sinh(\psi-\psi_k)$ (see
(\ref{vvz:8})). It means that $\psi(x)\to \psi_k$ for $k=1,\dots,n$. It is possible only if functions $a_k(x)$,
$k=1,\dots,n$, $\psi(x)$, and $E(x)$ are close to stepwise ones.

Let us introduce the set of intervals $(x_{k-1},x_k)\in[0,1]$ that are not overlapping. We assume that
\begin{eqnarray*}
&& a_k(x)\equiv a_{k}^{0}, \quad
   x\in(x_{k-1},x_{k})\quad
   k=1,\ldots ,n,
\end{eqnarray*}
\begin{eqnarray}
&& a_k(x)\equiv 0, \quad x \notin (x_{k-1},x_{k}), \quad
   k=1,\ldots ,n,
   \label{vvz:20}
\end{eqnarray}
where $a_k^0$ are some constants. Then a particular conclusion from (\ref{vvz:15}) at $K_w \to 0$ is:
\begin{eqnarray}
&& \psi(x) \equiv\psi_{k},\quad x\in(x_{k-1},x_{k}), \quad
k=1,\dots,n.
    \label{vvz:21}
\end{eqnarray}
Next we introduce the functions:
\begin{eqnarray*}
&&  \Phi_{k}(x)\equiv \frac{1}{\sigma(x)}
  \frac{1}{\varphi_{k}(\psi(x))}
  \frac{d\varphi_{k}(\psi(x))}{d\psi}, \quad
\end{eqnarray*}
\begin{eqnarray}
&&  F_{k}(x)\equiv \frac{1}{\varepsilon_{0}} \int\limits_{0}^{x}\Phi_{k}(s)\,ds, \quad k=1,\dots,n.
  \quad
  \label{vvz:22}
\end{eqnarray}
For an interval $(x_{i-1},x_{i})$ we get:
\begin{eqnarray}
&&  \Phi_{k}(x) \equiv
    \frac{e_k(\psi_i)}{\gamma_{i}a_{i}^{0}} =
    \Phi_{k}^{(i)},
   \label{vvz:23}
\end{eqnarray}
\begin{eqnarray*}
&&  \varepsilon_0 F_{k}(x)=\sum\limits_{s=1}^{i-1}
  \Phi_{k}^{(s)}(x_{s}-x_{s-1})+\Phi_{k}^{(j)}(x-x_{i-1}).
\end{eqnarray*}
The integrating of (\ref{vvz:17}) and taking into account notations (\ref{vvz:22}) yield:
\begin{eqnarray}
\label{c_k(x)}
   a_{k}(x)=a_{k}(0) e^{F_k(x)}.
   \label{vvz:24}
\end{eqnarray}
Using  (\ref{vvz:22}), (\ref{vvz:24}), and (\ref{vvz:21}), the equations  $a_k(x)=a_k^0$ for $x\in(x_{k-1},x_{k})$ give:
\begin{eqnarray}
a_{k}(0)=  a_{k}^{0}\exp\left(-\frac{1}{\varepsilon_{0}}
\sum\limits_{s=1}^{k-1}\Phi_{k}^{(s)}(x_{s}-x_{s-1}) \right).
   \label{vvz:25}
\end{eqnarray}
The values $a_k^0$ can be easily found from the boundary condition $m_k(1)=M_k$ with the help of (\ref{vvz:20}), (\ref{vvz:21}):
\begin{eqnarray}
   a_{k}^{0}=\frac{M_{k}}{(\delta_{k}+1)(x_{k}-x_{k-1})}, \quad k=1,\dots,n.
   \label{vvz:26}
\end{eqnarray}
Expressions (\ref{vvz:26}), (\ref{vvz:25}), (\ref{vvz:23}) represent the required initial approximation for $a_k(0)$, provided that the
lengths of intervals $(x_{k},x_{k-1})$, $k=1,\dots,n$ are given. The main asymptotic term at $K_w\to 0$ is given by (\ref{vvz:24}).

\section{Numerical Results}\label{i:5.0}

\begin{figure}[h]
  \centering
\includegraphics[scale=0.59]{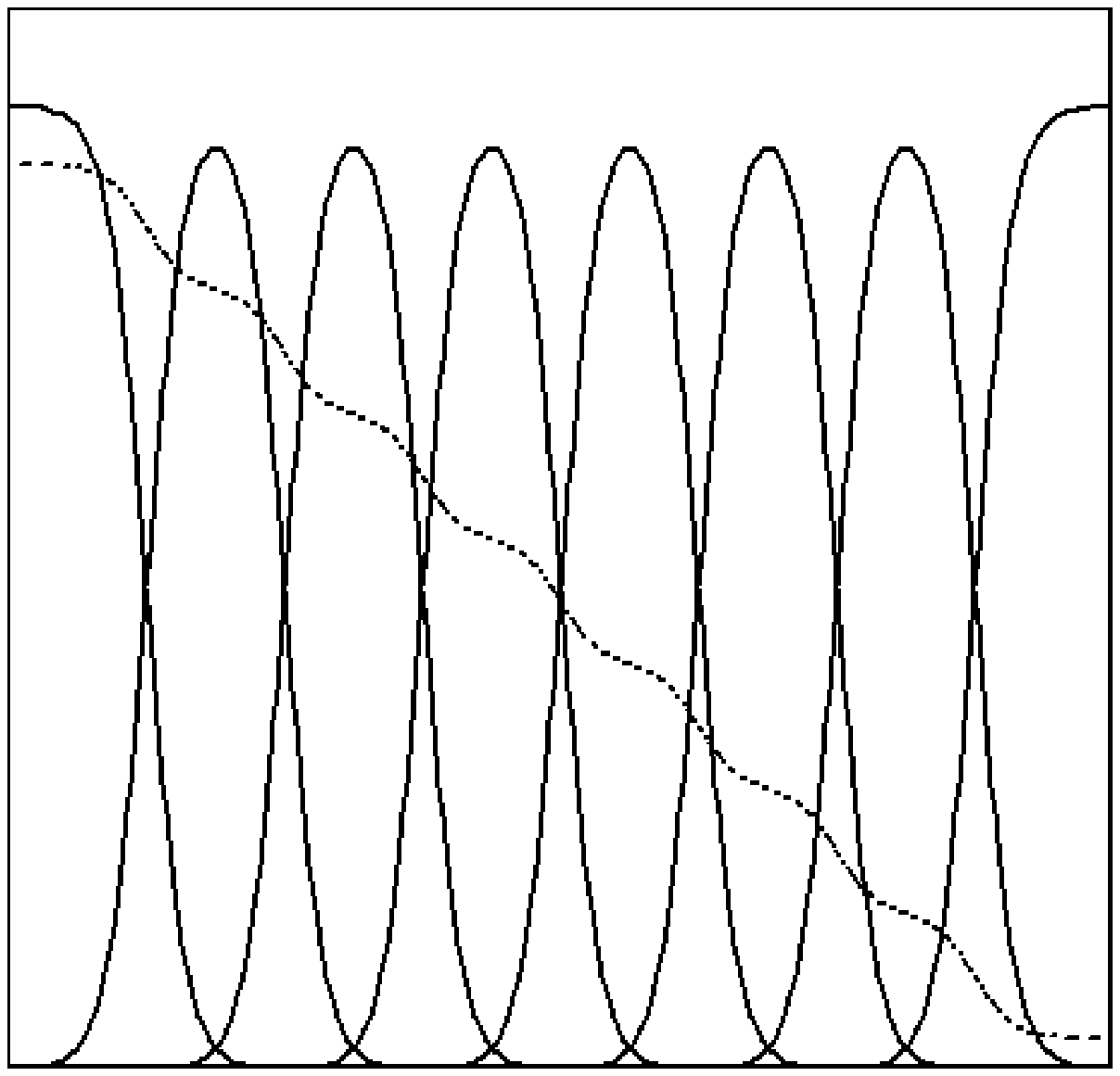}\,
\includegraphics[scale=0.59]{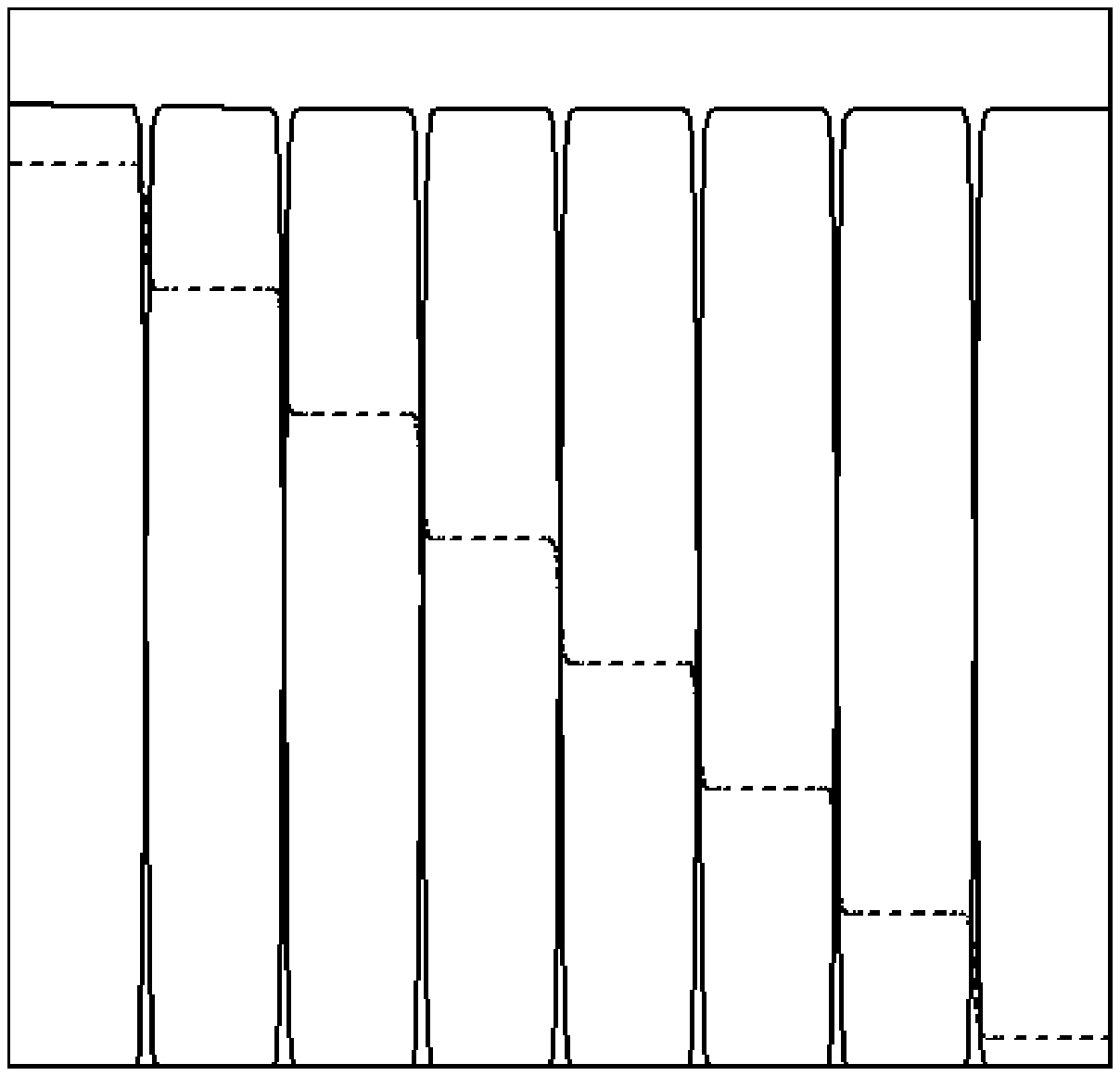}
  \caption{The concentration distribution  $a_k(x)$, $k=1,\dots,8$ and $\psi(x)$ (the dashed line)}
  \label{av_fig}
\end{figure}

The main difficulty in the use of formulas (\ref{vvz:25}), (\ref{vvz:26}) is the fact that the lengths of the
intervals $(x_{k-1},x_{k})$, $k=1,\dots,n$ are unknown, they can be determined by the next asymptotic terms. However
our numerical experiments have shown that (at least for $n>5$) it is sufficient to take their lengths equal:
$(x_{k}-x_{k-1})=1/n$. We have shown that asymptotic formulas and numerical results are in good agreement beginning
from $\varepsilon_0\approx 2\cdot 10^{-4}$ and $K_w\approx 10^{-4}$. In particular, the numerical results obtained by a
shooting method for the values of parameters $n=8$, $\delta_k=99$, $\psi_k=3.85-0.6k$, $M_k=160$,
$\varepsilon_0=5.25\cdot10^{-6}$, $\gamma_k=0.15$, $\psi_0=-0.274$, $\gamma_0=7.077$, $K_w=10^{-5}$ are given in
Table~\ref{av_tab:1} where for the sake of convenience we present the values $z_k=-10\,\varepsilon_0\ln a_{k}(0)$.
\begin{table}[h]
\begin{center}
\begin{tabular}{|c|c|c|c|c|c|c|c|}
  \hline
  $k$ &  2 & 3 & 4 & 5 & 6 & 7 & 8\\ \hline
  $z_{k}$
      &  0.0042 & 0.0138 & 0.0325 & 0.0662 & 0.1256 & 0.2264 & 0.3894\\ \hline
  $z^*_{k}$
      &  0.0039 & 0.0137 & 0.0324 & 0.0664 & 0.1262 & 0.2276 & 0.3914\\ \hline
\end{tabular}
\end{center}
\caption{The comparison between the numerical and asymptotic values} \label{av_tab:1}
\end{table}
The values $z_k^*$ in the second raw correspond to their asymptotic values. One should notice that the formulas
(\ref{vvz:25}), (\ref{vvz:26}) do not work well for $a_1(0)$, so for the obtaining of its asymptotic value one has to
use an additional procedure that are similar to Sect.~\ref{i:3.0} for $a_k(1)$.
Fig.~\ref{av_fig} shows the concentration distributions $a_k(x)$ and function $\psi(x)$ for two values
$\varepsilon_0=5.25\cdot10^{-5}$ (left) and $\varepsilon_0=5.25\cdot10^{-6}$ (right).
It is clearly seen that $a_k(x)$
are located successively on the $x$-axis in the interval $(x_{k-1},x_{k})$. The concentration maxima correspond to the
isoelectric points $\psi_k$. In a classical case when $\varepsilon_0$ is not too small,  function $\psi(x)$ is
monotonically decreasing and concentration distributions are close to Gaussian ones (the left frame). The decreasing of
$\varepsilon_0$ (or the increasing of current density) causes the `almost' stepwise distribution of $\psi(x)$
\emph{i.e.} $\psi(x)\approx \psi_k$ in the intervals $(x_{k-1},x_{k})$ and `almost' constant concentrations
$a_{k}(x)\approx a_{k}^0$ (the right frame).

\begin{acknowledgments}
This research is partially supported by EPSRC (research grants
GR/S96616/01, EP/D055261/1, and EP/D035635/1), by the Russian
Ministry of Education (programme `Development of the research
potential of the high school',  grants 2.1.1/6095 and 2.1.1/554),
and by Russian Foundation for Basic Research (grants 07-01-00389,
08-01-00895, and 07-01-92213 NCNIL). The authors are grateful to the
Department of Mathematics of the University of York for the
providing of excellent conditions for this
research.
\end{acknowledgments}

\end{document}